\documentclass[iop]{aastex}
\usepackage{natbib}
\usepackage{color}

\def\MO{$M_{\odot}$}

\def\MJ{$M_{J}$}
\def\Teff{$T_{\rm{eff}}$}
\def\Tcr{$T_{\rm{cr}}$}

\def\logg{log \textit{g}}
\def\Rj{$R_{J}$}

\def\CHf{$\mathrm{CH_4}$}

\def\AKARI{\textit{AKARI}}

\slugcomment{Submitted to ApJ, 2012 October 14. }

\shorttitle{Radii of Brown Dwarfs Obtained by \textit{AKARI}}
\shortauthors{Sorahana et al.}

\begin{document}


\title{On the Radii of Brown Dwarfs Measured with \textit{AKARI} Near-Infrared Spectroscopy}


\author{S. Sorahana$^{1,2}$, I. Yamamura$^{2}$ and H. Murakami$^{2}$}
\affil{$^{1}$Department of Astronomy, Graduate School of Science, The University of Tokyo,
Bunkyo-ku, \\
Tokyo 113-0033, Japan; sorahana@ir.isas.jaxa.jp \linebreak
${^2}$Department of Space Astronomy and Astrophysics, Institute of Space and Astronautical Science (ISAS),\\ Japan Aerospace Exploration Agency (JAXA), 
Sagamihara, Kanagawa 252-5210, Japan}


\begin{abstract}
We derive the radii of 16 brown dwarfs observed by {\textit{AKARI}} using their parallaxes and the ratios of observed to model fluxes. 
We find that the brown dwarf radius ranges between 0.64--1.13~$R_J$ with an average radius of 0.83~$R_J$.  
We find a trend in the relation between radii and {\Teff}; 
the radius is at a minimum at {\Teff}$\sim$1600~K, 
which corresponds to the spectral types of mid- to late-L.  
The result is interpreted by a combination of radius--mass and radius--age relations that are theoretically expected for brown dwarfs older than 10$^{8}$~yr.
\end{abstract}





\section{Introduction}
Brown dwarfs are generally defined as the objects that are too light to maintain hydrogen fusion in their cores. 
For the solar metallicity, the upper limit of the brown dwarf mass, i.e., the hydrogen-burning main sequence edge mass, is 0.08~{\MO}
(\citealt{Kumar_1963}; \citealt{Hayashi_1963}). 
The lower edge of mass range of brown dwarfs overlaps with mass of planets.  
They are distinguished by their formation process instead of mass. 
Brown dwarfs are born in the interstellar medium through processes similar to that form stars. 
Objects that are born in protoplanetary disks should be referred to as planets. 

The radii of brown dwarfs and planets in a context of their evolution are interesting subject. 
Radius of an exoplanet can be derived from photometric transit observations.
On the other hand, it is difficult to determine a radius of a brown dwarf observationally, 
and that has not been discussed so far except for the objects in binary system. 
According to the theoretical study by \citet{Burrows_2001} 
some brown dwarfs are actually smaller than Jupiter. 
The radii of brown dwarfs older than $10^{9}$~yrs remain roughly constant at 0.7--1.1~{\Rj} ({\Rj}: Jupiter's radius) for a broad mass range from 0.3 to 70~{\MJ}. 
Interestingly, more massive dwarfs have smaller radii.
This fact is a consequence of the competition between two equations of state; Coulomb and electron degeneracy effects. 
Since the objects held by Coulomb effect would only set a fixed density, the relation of radius and mass follows $r \propto M^{1/3}$.  
On the other hand, electron degeneracy results in $r \propto M^{-1/3}$. 
Competition between two effects leads to the roughly constant radius--mass relation at $\ge 10^9$ yrs. 
\citet{Burgasser_phd} evaluated the expected radius of an observed brown dwarf by performing a Monte Carlo simulation adopting the evolutionary models by \citet{Burrows_1997} with assumption of a constant birth rate and  mass function $\frac{dN}{dM}\propto M^{-1}$. 
He found that the most likely radius is 0.9$\pm$0.1~{\Rj}. 

Effective temperature ({\Teff}) of a brown dwarf has often been estimated empirically by assuming its radius. 
\citet{Vrba_2004} used a constant radius of $r$ = 0.9~{\Rj}  following \citet{Burgasser_phd} to estimate the empirical {\Teff} of 40 L and T dwarfs. 
They estimated {\Teff} from the equation $L=4\pi R^2\sigma T_{\rm{eff}}^4$ 
(\citealt{Drilling_2000}), where luminosity is derived from $K$-band flux by bolometric corrections. 
They reported that their derived {\Teff} for the early-L dwarfs, about 2400--2500~K, are warmer by about 200--300~K than some earlier estimates by fitting the spectral energy distributions with the synthetic spectra (e.g., \citealt{Leggett_2001}). 
They argued that it is caused by different photometry database and slightly different assumptions of brown dwarf radii.

Yamamura et al. (2010) analyzed 2.5--5.0~$\mu$m spectra of six brown dwarfs taken by the Japanese infrared astronomical satellite {\AKARI} \citep{Murakami_2007}. 
They found that {\Teff} derived by the Unified Cloudy Model (UCM; \citealt{Tsuji_2002,Tsuji_2005}) fitting to the observed spectra are lower by typically 200~K from the empirical {\Teff} estimated by \citet{Vrba_2004}. 
They also reported that the radii derived from the {\AKARI} observations have an average value of 0.81~{\Rj}, 
and distribute in a wider range between 0.68--1.18~{\Rj} than that expected from \citet{Burrows_2001}. 
They argued that the radii of brown dwarfs should not be represented by a single mean value. 

In this paper, we discuss the radii of an extended sample of brown dwarfs determined by the {\AKARI} observations in detail.  
First, we introduce our objects in Section~\ref{sample}, 
and outline our fitting procedure briefly in Section~\ref{bfm} \footnote{The details of data reduction, calibration and fitting process are described in \citet{Sorahana_2012}.}.  
We describe a method of deriving radius of brown dwarfs and uncertainty of the radius in Section~\ref{radius4}. 
We present the resulted radii of 16 observed brown dwarfs in Section~\ref{result}, 
and discuss the relation between radius and {\Teff} and their evolutionary status in Section~\ref{discussion}. 

\section{The {\AKARI} Sample}
\label{sample}
Twenty seven brown dwarfs including 16 L dwarfs and 11 T dwarfs were observed by {\textit{AKARI}}. 
Our targets selected by their expected fluxes to be bright enough
for Infrared Camera (IRC) onboard the {\AKARI} to provide high-quality spectra within the reasonable amount 
of exposures and their spectral types to sample various types from L to T. 
We obtained good quality spectra of 16 brown dwarfs included 11 L and 5 T dwarfs (Table~\ref{radmasslist}). 

These 16 objects are nearby and bright, thus they are generally well studied.
There are three binaries in this sample, GJ~1001B, 2MASS~J1523+3014 and Gl~570D. 
2MASS~J1523+3014 is as known as Gl~584C and is a companion of Gl~584AB, which is a G dwarf double. 
2MASS~J1523+3014 is widely (194$''$) separated from Gl~584AB (\citealt{Kirkpatrick_2000}). 
Gl~570D is a companion of Gl~570ABC triple system, and is also located 258$''$ from Gl~570ABC (\citealt{Burgasser_2000}). 
In our observation, the target source was placed in the $1\times1$ arcmin$^2$ aperture of the {\AKARI}/IRC instrument.  
Therefore, 2MASS J1523+3014 and Gl~570D were observed without confusion from their primary stars.
On the other hand, GJ~1001B is located only 18.2$''$ from the primary M dwarf GJ~1001A (\citealt{Goldman_1999}) 
and the spectrum of GJ~1001B was contaminated by a shoulder of an intense signal from GJ~1001A. 
We evaluated and subtracted the signal from GJ~1001A at the position of GJ~1001B (see \citealt{Sorahana_2012} for detail).
\citet{Golimowski_2004b} found that GJ~1001B also has a companion GJ~1001C separated by 0.087$''$. 
The two dwarfs are of the similar spectral types.  
To derive the radius of GJ~1001B (or C; hereafter GJ~1001B),  
we assume that the luminosity of each dwarf is a half of the observed one.
All other objects in our sample are believed to be single sources.

\begin{deluxetable}{lcrrcc}
\tabletypesize{\scriptsize}
  \tablecaption{Sixteen Brown Dwarfs observed by {\AKARI} \label{radmasslist}}
\tablewidth{0pt}
\tablehead{
\colhead{Object Name}  &\colhead{Sp. Type} &\colhead{{\Teff}[K]}&  \colhead{Parallax(error)[mas]}  &  \colhead{Binary} &  \colhead{References} 
}
\startdata
2MASS~J14392836+1929149&L1 &2100&69.6(0.5)&No&1, a\\
2MASS~J00361617+1821104 & L4 &2000&114.2(0.8)&No&2, a\\
2MASS~J22244381--0158521 & L4.5 &1800&85.0(1.5)&No&1, b\\
GJ~1001B & L5 &1800&76.9(4.0)&Yes&1, c\\
SDSS~J144600.60+002452.0  &L5    &1800&45.5(3.3)&No&2, b\\
SDSS~J053951.99--005902.0 &L5   &1800&76.1(2.2)&No&2, b\\
2MASS~J15074769--1627386&L5  &1800&136.4(0.6)&No&1, a\\
2MASS~J08251968+2115521 & L6 &1500&95.6(1.8)&No&2, b\\
2MASS~J16322911+1904407 &L7.5  &1500&63.6(3.3)&No&2, b\\
2MASS~J15232263+3014562 & L8 &1600&57.3(3.3)&Yes&2, b\\
SDSS~J083008.12+482847.4 & L9 &1600&76.4(3.4)&No&2, b\\
SDSS J125453.90--012247.4 & T2 &1400&75.7(2.9)&No&2, b\\
SIMP J013656.5+093347.3 &T2.5  &1400&{\bf 6.4(0.3)}&No&3, d\\
2MASS J05591914--1404488 &T4.5  &1200&95.5(1.4)&No&2, b\\
Gliese~570D & T8 &700&169.3(1.7)&Yes&2, a\\
2MASS~J04151954--0935066 &T8  &700&174.3(2.8)&No&3, b    
\enddata
\tablecomments{Reference of spectral type (1) \citet{Kirkpatrick_2000}, 
(2) \citet{Geballe_2002}, 
(3) \citet{Burgasser_2006}.\\
The parallaxes are referred from (a) \citet{Dahn_2002}, (b) \citet{Vrba_2004}, (c) \citet{Henry_2006}, 
and (d) \citet{Artigau_2006}. \\
The number given for SIMP~J0136+0933 is a photometric distance [pc] estimated by comparing the spectral energy distribution with known brown dwarfs of similar spectral types. 
}
 \end{deluxetable}

\section{Best Fit Model Derived from {\AKARI} Near-Infrared Spectra}
\label{bfm}
\begin{figure}
\begin{center}
   \plotone{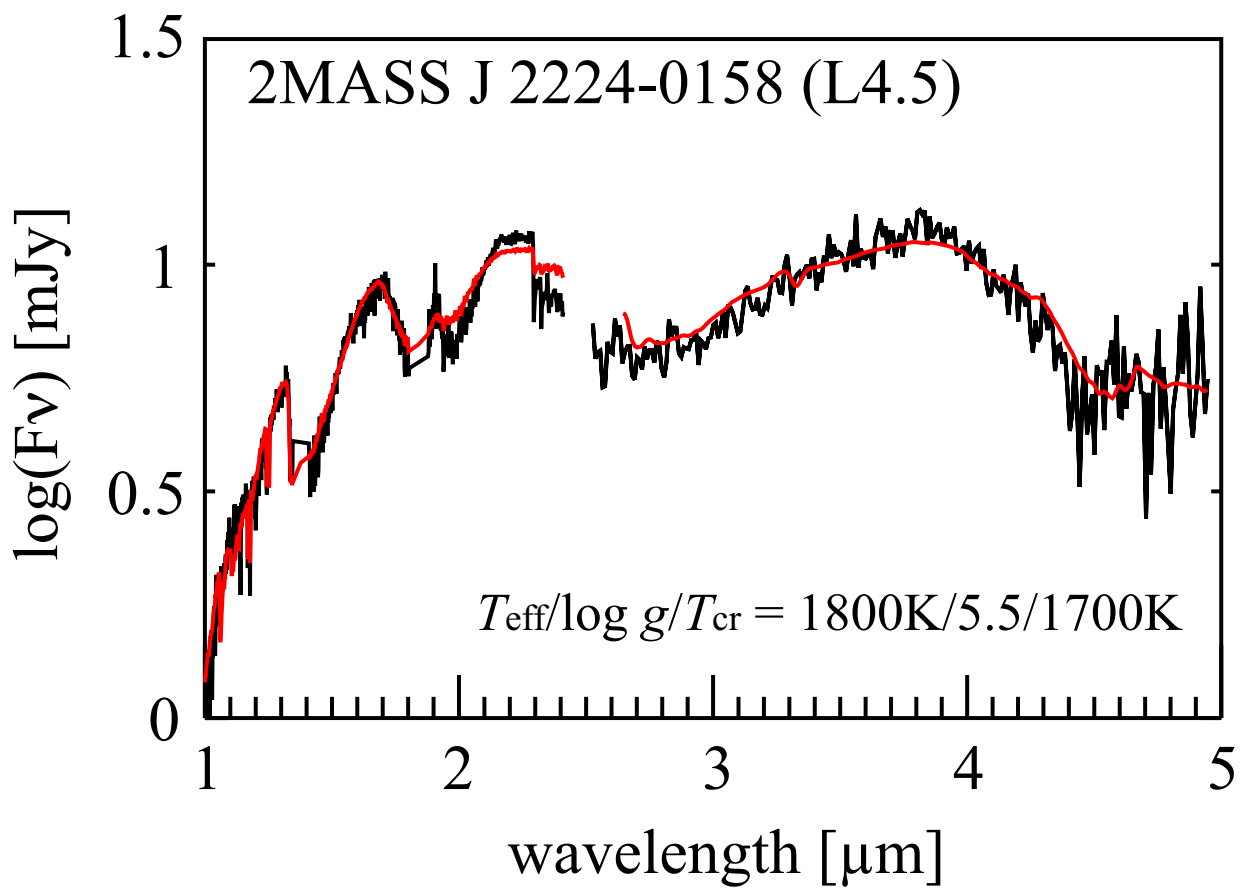}
\end{center}
\caption{The observed and the best fit model spectra for 2MASS~J2224--0158 (L4.5).
The black line is the observed data and the red line is the best-fit model spectra.
The spectrum between 2.5 and 5.0 $\mu$m is taken by {\AKARI}
and that between 1.0 and 2.5 $\mu$m is taken by IRTF/SpeX. 
The parameters of the best fit model are given in the legend of the figure.
} 
\label{examfit}
\end{figure}

We derived physical parameters of brown dwarfs, namely effective temperature {\Teff}, surface gravity {\logg} and critical temperature {\Tcr} by model fitting with UCM. 
{\Tcr} is a temperature below which the dust disappears by sedimentation or other unknown mechanism, 
and given as an additional parameter in UCM, 
i.e., the dust would exist in the layer with {\Tcr}$<T < T_{cond}$.
{\Tcr} is not predictable by any physical theory at present and is required to be determined from observations empirically.
We mainly use {\AKARI} spectra in the range of 2.5--4.15~$\mu$m (not to 5.0 $\mu$m because the current
model does not explain the observed spectra beyond 4.15~$\mu$m. See \citealt{Yamamura_2010}). 
We follow \citet{Cushing_2008} and evaluate the goodness of the model fitting by the statistic $G_k$ defined as
\begin{equation}
\label{gk}
G_{k} = \frac{1}{n-m}\sum_{i=1}^n \omega_{i}  \left( \frac{f_{i} - C_{k}F_{k,i}}{\sigma_{i}} \right)^2,
\end{equation}
where $n$ is the number of data points; $m$ is degree of freedom (this case $m=3$); $\omega_{i}$ is the weight for the $i$-th wavelength points 
(we give the equal weight as $\omega_{i}$ = 1 for all data points because of no bias within each observed spectrum); 
$f_{i}$ and $F_{k, i}$ are the flux densities of the observed data and $k$-th model, respectively; 
$\sigma_{i}$ are the errors in the observed flux densities 
and $C_{k}$ is the scaling factor given by 

\begin{equation}
\label{scalingfactor}
C_{k} = \frac{\sum \omega_{i} f_{i} F_{k,i}/\sigma_{i}^2}{\sum \omega_{i} {F_{k,i}}^2/{\sigma_{i}}^2}.
\end{equation}
$G_{k}$ is equivalent to reduced $\chi ^{2}$, since we adopt $\omega_{i}$ = 1 in our analysis. 
It is difficult to determine a unique best-fit model for each object because of relatively large error associated with the
{\AKARI} spectra.
Thus we also use the shorter wavelength spectra (IRTF/SpeX and UKIRT/CGS4
data) to complete our analysis.
Details of fitting evaluation are described in \citet{Sorahana_2012}. 
We show an example of model fitting in Figure~\ref{examfit}. 
This object, 2MASS~J2224--0158 is a relatively warm object of the spectral type L4.5. 

\section{Radius of Brown Dwarfs}
\label{radius4}
\subsection{Derivation of the Radius}
\label{drbd}
The ratio of model flux $F_{\nu}$ [erg/s/cm$^{2}$/Hz], which is for a unit radius at a unit distance, to observed flux $f_{\nu}$ [mJy] is written as 
\begin{equation}
\label{eq1}
\log(F_{\nu}/f_{\nu})=-2\log(r/D)-26.497, 
\end{equation}
where $r$ is the radius and $D$ is the distance.
$D$ of each object has been calculated from its parallax given in the 4th-column of Table~\ref{radmasslist}, except for SIMP~J0136+0933 for which the value is a spectrophotometric estimate. 
Distance of SIMP~J0136+0933 is estimated by comparing the flux with known brown dwarfs of similar spectral types (\citealt{Artigau_2006}). 
The model flux $F_{\nu}$ is from the UCM best fit model spectra given by \citet{Sorahana_2012}. 
Equation (\ref{eq1}) is written with $C$ of the best fit model as, 
\begin{equation}
\frac{C}{10^{26.497}} = \left(\frac{r}{D} \right)^2.
\label{r}
\end{equation}

\subsection{Uncertainty of the Radius}
\label{uncertaintyc}
The radius in this analysis relies on the distance $D$ and the flux scaling factor $C$ defined in Equation (\ref{scalingfactor}). 
The uncertainty of $D$ is estimated from the parallax error shown in Table~\ref{radmasslist}.  
The maximum error of $D$ is $\sim$7~\%. 
Some brown dwarfs have multiple parallax measurements and the results are consistent to each other within the error. 
The uncertainty of $C$ depends on the absolute flux calibration of {\AKARI} spectra and the goodness of the model fitting.
The former is evaluated to be about 10~\% (\citealt{Ohyama_2007}), 
and was also validated with $L'$ photometry data in \citet{Sorahana_2012}. 
The latter error is different for each object. 
Model spectra do not yet perfectly reproduce the observed spectra because of  
for example an incomplete {\CHf} line list and unknown properties of dust in the photosphere not yet incorporated into the model (\citealt{Yamamura_2010}). 
Hence it is hard to find a unique best-fit model.
\citet{Sorahana_2012} discussed the uncertainty of the best-fit model parameters. 
The uncertainty should be no better than a half of the grid spacing (100~K for {\Teff} and {\Tcr} and 0.5~dex for {\logg}). 
To estimate the uncertainty we change each of {\Teff}, {\logg}, and {\Tcr} by one grid from the best-fit value, 
and search for the $``$restricted best$"$ model by changing other two parameters following the same manner through fitting evaluation. 
If we do not find any models satisfying $G_{min} \le G_k < G_{min} + 1$ (here, $G_{min}$ is taken from all parameter space in model fitting with {\AKARI} data only), then the uncertainty of the parameter should be smaller than the grid spacing. 
In order to estimate the uncertainty of $C$, 
we evaluate the ratio of $C'$ for the $``$restricted best$"$ model to $C$ for the best model.  
The maximum deviation of $C'/C$ from 1.0 is $\sim$30~\%. 
The resulted overall error of derived radius is listed in Table \ref{radmasslist2}. 
Uncertainty of radius ranges between 5~\% and 16~\%. 
The uncertainty is the largest for 2MASS~J0559--1404 because of the large $C$ error.

\section{Results}
\label{result}
The result is listed in Table~\ref{radmasslist2}.  
The average radius of 16 brown dwarf samples is 0.83~{\Rj} with a standard deviation of 0.14~{\Rj}. 
The average is slightly smaller than the value given by \citet{Burgasser_phd}, 
but is consistent within the uncertainty.
The derived radii  
ranges from 0.64 to 1.13~{\Rj}. 
Figure~\ref{rad2} shows the relation between {\Teff} and radii at different masses from 11~{\MJ} to 0.085~{\MO} predicted by evolutionary model of \citet{Burrows_2001}, 
and our results are over-plotted with uncertainties. 
Our results are consistent with theoretical predictions for early-L and T dwarfs. 

From this figure we can discuss mass and age of the objects.
Both masses and ages of 2MASS~J0036+1821 and 2MASS~J2224--0158 are similar to each other, though their {\Teff} are different. 
Ages of 2MASS~J1439+1929 and 2MASS~J0559--1404 are close (relatively younger than other objects) to each other, but their masses are quite different. 
Likewise, SDSS~J1254--0122 and 2MASS~J0415--0935 are in similar ages, but with different masses. 
On the other hand, masses of 2MASS~J0559--1404 and Gl~570D are close to each other, but ages are not;  
2MASS~J0559--1404 is younger than that of Gl~570D. 
The age of Gl~570D is estimated by \citet{Kirkpatrick_2001} as 2--10~Gyr, and is consistent with our result within its uncertainty.

Our sample includes objects with very small radii. 
Radii of GJ~1001B, 2MASS~J1523+3014, SDSS~J0830+4828, SDSS~1446+0024 and 2MASS~J1507--1624 are 0.64, 0.65, 0.66, 0.74 and 0.79, respectively. 
They are out of range of the theoretical predictions.
Interestingly those small radii objects are all around {\Teff}$\sim$1600~K, 
i.e., the radii of our brown dwarfs hold minimum at mid- to late-L, 
though we apply some assumptions for GJ~1001B through the data analysis.

\section{Discussion}
\label{discussion}
We find that the radii of mid- to late-L objects with {\Teff}$\sim$1600~K are smaller than those of theoretical predictions in absolute scale,  
and those of early-L and T dwarfs in relative scale. 
In this section, we discuss these results. 

\subsection{Validity of Our Result}
\label{quan}
First, we verify the absolute values of our resultant radii. 
A fair comparison with the theoretical prediction needs a mass and an age of each brown dwarf. 
However, it is difficult to measure the accurate mass and age of each object observationally unless the objects are members of a cluster or a binary.
If the object is unresolved binary for {\AKARI}, their radius should be even smaller. 
Mass can be evaluated from the surface gravity. 
However, the uncertainty of mass of each object is very large because of the large uncertainty in surface gravity derived by the model fitting (\citealt{Sorahana_2012}).

We can justify our result in another way. 
If we assume that the radius of 2MASS~J1523--3014 is the mean value 0.83~{\Rj}, the flux levels should be 1.6--1.7 times higher. 
This is unreasonably lager than the error in the flux calibration, and we regard that the radii of mid- to late-L dwarfs should be as small as 0.7~{\Rj}.
It is noted that the spectra of mid- to late-L dwarfs are most affected by the dust in the atmosphere. 
The incompleteness in the atmosphere model could draw a systematic error in the flux level. 
However, we do not see any particular jumps in the flux levels of the models along the changes of the model parameters, 
indicating that 60--70 \% changes in the flux level should not be accounted by the model.

Recently, \citet{Burrows_2011} discuss metallicity dependence of brown dwarf radii. 
In fact we argue that our {\AKARI} sample may have metallicity variation (\citealt{Tsuji_2011}; Sorahana et al. in prep.).
However, the expected change of radius is as much as several per cent still much smaller than the discrepancy we found here. 

We conclude that the radii derived from our observed data are regarded to be real
and admit there exists a serious deviation between observation and evolutional model.

\subsection{Radius Inversion}
\label{qual}
\begin{figure}
\begin{center}
    \plotone{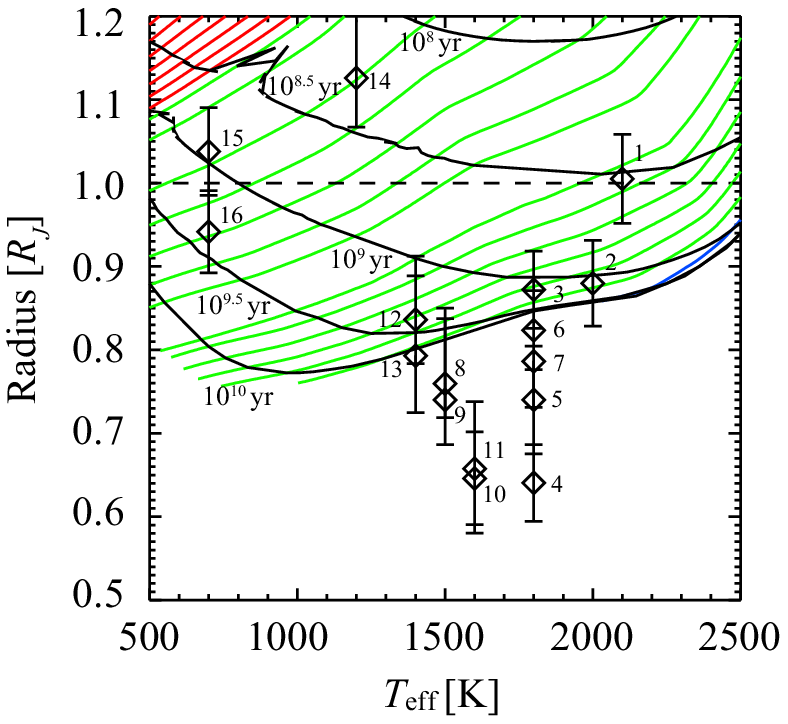}
\end{center}
\caption{The radii of {\AKARI} observed brown dwarfs are plotted on theoretical prediction of the relation between {\Teff} and radii at different masses from 11~{\MJ} to 0.085~{\MO}. 
Red are less massive objects with 11, 12, 13 and  15~{\MJ}, blue are massive objects 0.08 and 0.085~{\MO} and green are intermediate mass objects with 0.02, 0.025, 0.03, 0.035, 0.04, 0.045, 0.05, 0.055, 0.06, 0.065, 0.07 and 0.075~{\MO}. 
The numerical data of evolutionary model are provided by Adam Burrows (in private communication). 
The numbers are object index in Table~\ref{radmasslist2}. 
} 
\label{rad2}
\end{figure}

Our {\AKARI} sample is selected by their spectral types to sample various types from L to T, 
and may be biased to some extent. 
They locate very close to our sun (5--25 pc),  
thus we expect that they are as old as the sun ($\sim$10$^{9.7}$~yrs). 
Ages of three binaries in our sample, GJ~1001B, 2MASS~J1523+3014 and Gl~570D, are determined by \citet{Kirkpatrick_2001} 
to be 1~Gyr, 1-2.5~Gyr and 2-10~Gyr, respectively.
The ages of our objects can also be evaluated from Figure~\ref{rad2}, as we discussed in this section. 
Those of mid- to late-L dwarfs are out of the theoretical evolutionary tracks but they are likely older than 10$^8$~yr.
Our results indicate older ages for GJ~1001B and 2MASS~J1523+3014. 
While theory allows a wide variety of mass for T dwarfs, 
all L dwarfs at near solar age are massive objects (See Figure~8 of \citealt{Burrows_2001}). 
In other words, it is highly possible that L dwarfs of our sample are more massive than our T dwarfs. 
On the other hand, Figure~\ref{rad2} tells that early-L dwarfs in our sample (e.g., 2MASS J439+1929, 2MASS J0036+1821) are relatively younger and less massive than mid- to late-L objects.

\begin{figure}
\begin{center}
   \plotone{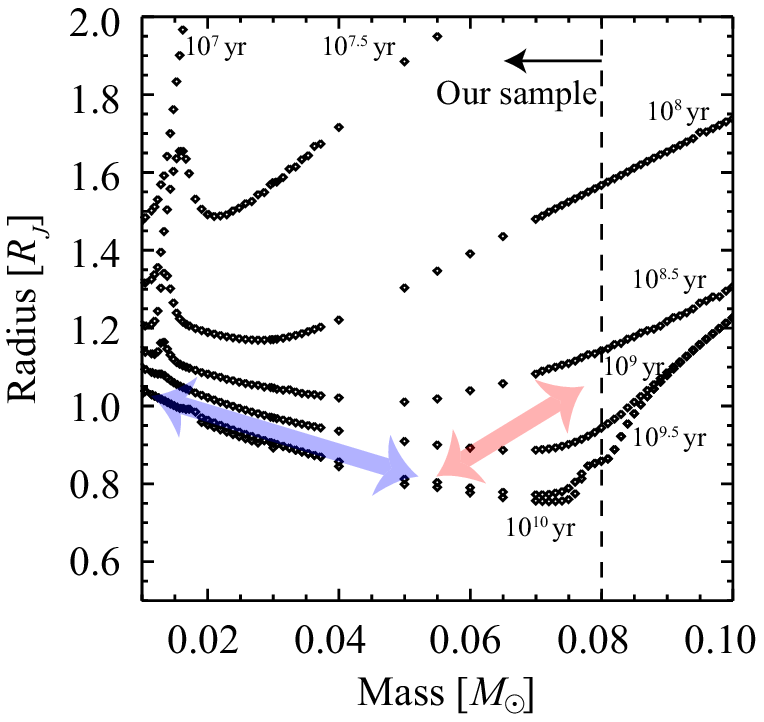}
\end{center}
\caption{Theoretical prediction of the relation between radius and mass of brown dwarfs at different ages. 
The radii of relatively light and young objects are small and rather stable throughout their lifetimes. 
On the other hand, the radii of massive dwarfs $\ge$~0.06~{\MO} are large when they are young, 
but ultimate radii are smaller than those of less massive objects.  
The radii of early-type objects in our sample tend to depend on their age (shown by red arrow), 
while those of late-type objects tend to depend more on their mass than their age (shown by blue arrow). 
The data are provided by Adam Burrows (in private communication). } 
\label{radage}
\end{figure}

Brown dwarfs shrink slowly as they evolve. 
As previously noted, the radii of brown dwarfs do not follow a monotonic function of mass, in particular after the age of $10^{7.5}$~yrs. 
According to \citet{Burrows_2001}, 
the radius of a less massive object is already small at its formation and change of radius during its lifetime is also small. 
On the other hand, the radius of a massive object is large when it is young,  
but become smaller than the lighter objects beyond $\sim 10^{8}$~yrs, 
because the electron degeneracy effect overcomes the coulomb effect. 
This inversion of radius takes place continuously in the mass range between 0.3 and 70~{\MJ}. 
The relation of radius and mass with age is shown in Figure~\ref{radage}. 
The data are provided from Adam Burrows (in priv. comm.).  
We see a depression on the curves for the dwarfs older than 10$^{8}$~yrs, 
which indicates the radius inversion by the degeneracy effect. 
A sharp bump at $\sim$0.015~{\MO} is due to a deuterium burning.
The horizontal axes of Figure~\ref{rad2} and \ref{radage} are not simply equivalent, 
because effective temperature {\Teff} (spectral type) of a brown dwarf depends on both mass and age. 
According to the discussions above, the trend in our {\AKARI} sample can be explained by a combination of both effects. 
Early types that the effective temperature ranges between 1500 and 2100~K tend to depend on their age as shown in Figure~\ref{rad2}. 
We show this trend by overlaying a red arrow in Figure~\ref{radage}. 
On the other hand, late-types with {\Teff} lower than 1500~K is more mass dependent than age.
The trend is described by a blue arrow in Figure~\ref{radage}. 
Our result of the relation of radii against their {\Teff} implies that inversion of radius predicted by theory is actually taking place.

\begin{deluxetable}{lccc}
\tabletypesize{\scriptsize}
  \tablecaption{The radius of {\AKARI} objects\label{radmasslist2}}
\tablewidth{0pt}
\tablehead{
\colhead{Object Name} &\colhead{Sp. Type} &  \colhead{Radius(error) [{\Rj}]} &  \colhead{Number in Figure~\ref{rad2}} 
}
\startdata
2MASS J1439+1929 &L1 &1.01(0.05)&1\\
2MASS J0036+1821 & L4 &0.88(0.05)&2\\
2MASS J2224--0158& L4.5 &0.87(0.05)&3\\
GJ~1001B & L5 &0.64(0.05)&4\\
SDSS J1446+0024 &L5    &0.74(0.06)&5\\
SDSS J0539--0059 &L5   &0.82(0.05)&6\\
2MASS J1507--1627&L5  &0.79(0.06)&7\\
2MASS J0825+2115& L6 &0.76(0.09)&8\\
2MASS~J1632+1904 &L7.5  &0.74(0.10)&9\\
2MASS~J1523+3014 & L8 &0.65(0.09)&10\\
SDSS~J0830+4828 & L9 &0.66(0.08)&11\\
SDSS~J1254--0122 & T2&0.84(0.05)&12\\
SIMP~J0136+0933 &T2.5  & 0.80(0.12)&13\\
2MASS~J0559--1404 &T4.5  & 1.13(0.18)&14\\
Gl~570D & T8 &1.04(0.05)&15\\
2MASS~J0415--0935 &T8  &0.94(0.05)&16   
\enddata
 \end{deluxetable}

\section{Conclusion}
We derive the radii of 16 brown dwarfs observed by {\AKARI} by comparing model flux and observed flux at given distances. 
The resulted radii ranges in 0.64--1.13~{\Rj}. 
The average radius is 0.83~{\Rj} with a standard deviation of 0.14~{\Rj}.
Our results are consistent with the theoretical radii calculated by \citet{Burrows_2001} within the error for early-L and late-T dwarfs, 
however, a large discrepancy is found for some mid- and late-L dwarfs. 
We find that the radii of mid- to late-L dwarfs are smaller than theoretical prediction. 
We verify that our estimates of radius is reasonable,  
and conclude that there are deviation between observation and evolutionary model. 
We can not find any reason of this deviation 
and leave it to the future studies. 
We also find that the radii reach a minimum for the mid- to late-L dwarfs.
Theory predicts that there is an inversion in the radius--mass relation in the brown dwarfs older than 10$^8$~yrs. 
Our results confirm that this theoretical prediction of radius inversion actually present. 

\vspace{0.5cm}
We thank to the anonymous referee for critical reading of our article and constructive suggestions.
This research is based on observations with {\AKARI}, a JAXA project with the participation of ESA.  
We thank to Prof. Takashi Tsuji for his kind permission to access the UCM and helpful suggestions. 
Prof. Adam Burrows kindly provide us the numerical data and warm encouragement.
We acknowledge JSPS (PI: S. Sorahana) and JSPS/KAKENHI(c) No. 22540260 (PI: I. Yamamura).




\bibliography{sora0404}

\end{document}